\documentclass[aps,prl]{revtex4}
\def\ee{\end{equation}}
\def\bea{\begin{eqnarray}}

\def\ket#1{| #1\rangle}

\def\Prob{{\rm Prob}}
\begin{document}

\title{Lorentzian Quantum Reality: Postulates and Toy Models}

\author{Adrian \surname{Kent}}
\affiliation{Centre for Quantum Information and Foundations, DAMTP, Centre for
  Mathematical Sciences, University of Cambridge, Wilberforce Road,
  Cambridge, CB3 0WA, U.K.}
\affiliation{Perimeter Institute for Theoretical Physics, 31 Caroline Street North, Waterloo, ON N2L 2Y5, Canada.}
\email{A.P.A.Kent@damtp.cam.ac.uk} 

\date{November 2014} 

\begin{abstract}
We describe postulates for a novel realist version of  
relativistic quantum theory or quantum field theory in Minkowski
space and other background spacetimes, and illustrate 
their application with toy models. 

\end{abstract}
\maketitle
  
\section{Introduction}

In previous papers \cite{akrealworld,akreality}, we
described ideas aimed at defining realist and
Lorentz covariant versions of relativistic quantum
theory or quantum field theory in Minkowski space.
The definitions also extend to other background
spacetimes with appropriate asymptotic properties
in the far future.
This paper introduces new ideas into this programme,
including a new and different proposal for the definition of beables.

The realist version of quantum theory proposed here
is defined by postulates that involve only expressions
obtained from the predictions that standard quantum
theory would make for various
specified measurements on the unitarily evolved
quantum state.   This makes it relatively easy
to illustrate its implications in
toy models.   The proposal relies on the
existence of particles, or
wave packets, or field perturbations, that
travel at light speed.   In this sense it
is intrinsically relativistic.  It is motivated
by the fact that quantum electrodynamics suggests it is 
sensible -- albeit not rigorously justifiable -- in some
circumstances to think of photons as individual particles 
that do indeed travel at
essentially light speed in the vacuum.   
Most approaches to unifying quantum theory and gravity suggest 
similar intuitions about gravitons, although of course
the direct evidence for quantum
gravity is thin.    

In a certain sense these light-speed particles (packets, field
perturbations) play the role of an environment, in that we
infer the actual state of a physical system in a given region
of space-time from the final state of some of these particles.
However, we do not need to postulate any formal separation between
``system'' and ``environment'' particles.   Our postulates treat
all types of particle, and all subsystems of the universe, on 
an equal footing. 

The proposal is
most simply illustrated in models that include
photons travelling along light ray segments along
with slower-moving systems made up of massive particles,
with particle number and photon number both conserved.     
Although 
models of this type can be good approximations in the
appropriate regime, they are not rigorously describable
by relativistic quantum field theory, since particle
numbers are not conserved in relativistic quantum field theory
and since we anyway have no rigorous formulation of non-trivial
relativistic field theories.   Nor are they rigorously described
by non-relativistic quantum mechanics, since we need both
that photon wave packets travel at the speed of light and that
massive particle wave packets travel at slower than light speed.
Nonetheless, they incorporate
standard intuitions about the behaviour of 
an essentially non-relativistic physical system
interacting with photons in the environment. 

Also, because the proposal requires distinguishing
between future data that could arise inside or 
outside the future light cone of a given point
in space-time, it does not behave well in 
the non-relativistic limit.    However, one
can try to define non-relativistic postulates
that produce approximately similar pictures of reality, even
if they rely on somewhat different assumptions.  
In another paper \cite{nonrel}, we will consider some postulates of this
type, which are also of some independent interest.  

\section{The reality problem for relativistic quantum theory} 

As noted in Ref. \cite{akreality}, since relativistic quantum field
theory is not itself presently based on a rigorous foundation, we
cannot hope to give a fully rigorous definition of a realist version
of that theory.  However, we {\it can} aim to give a conceptually
coherent description of reality within relativistic quantum field
theory, even if it involves quantities that presently have no rigorous
mathematical definition.  We can also illustrate the implications of
our postulates in toy models that incorporate intuitions from
relativistic quantum field theory but do not assume the full theory.
These are the main aims of this paper.
We put to one side here the precise proposals made in
Ref. \cite{akreality} in favour of the new simpler formulation given below.

We now suppose that the initial state $\ket{\psi_0}$ is given on some
spacelike hypersurface $S_0$, and that some relativistic unitary
evolution law is given.  The Tomonaga-Schwinger formalism allows us to
define formally the evolved state $\ket{\psi_S}$ on any hypersurface
$S$ in the future of $S_0$ via a unitary operator $U_{S_0 S}$.  These
future hypersurfaces $S$ play the same role here as in
\cite{akreality}.  That is, we initially give definitions for some
fixed hypersurface $S$ in the far future of $S$, but have in mind
ultimately taking the limit in which $S$ tends to future infinity.
For this asymptotic procedure to work, we have to assume that the
relevant limits are well-defined.  We examine this assumption in the
toy models we consider.

For any given hypersurface $S$ in the future of the initial
hypersurface $S_0$, we consider the effect of joint measurements of
the local mass-energy density operators $T_S (x) = T_{\mu \nu} (x)
n_{\mu} n_{\nu}$ carried out at each point $x \in S$, where $n_{\mu}$
is the forward-pointing timelike unit $4$-vector orthogonal to the
tangent plane of $S$ at $x$.  This gives us a probability distribution
on possible mass-energy distributions $t_S (x)$ on $S$.  In a universe
in which physics starts on $S_0$ and ends on $S$, our picture of
reality is that one $t_S (x) $ is randomly selected, from the Born
rule probability distribution.  In other words, there is a randomly
selected final boundary condition on $S$, which is defined
mathematically in the same way that it would be if $T_S (x) $ were
actually measured on $S$.  However, we treat this as simply as a
mathematical algorithm.  We do not suppose that a physical measurement actually
takes place on $S$, or anywhere else.  Our aim, instead, is to give a
mathematical description of reality applicable to closed quantum
systems, for which there are no external observers able to carry out
measurements.

To give a description of reality between $S_0$ and $S$, we use the
initial state on $S_0$, the randomly chosen final outcome data $t_S
(x)$ on $S$, and the unitary evolution law arising from the quantum
dynamics.  We use these data to calculate generalised expectation
values of the stress-energy tensor $\langle T_{\mu \nu} (y) \rangle$,
at each point $y$ between $S_0$ and $S$.  However, these expectation
values are not obtained in the familiar way by simply considering a
unitary evolution of the initial state to some spacelike hypersurface
through $y$.  Instead, we use the following recipe.

\subsection{Description of reality using generalized stress-energy tensor expectation values as beables}

We wish to define a generalized expectation value for the
stress-energy tensor at a point $y$ between $S_0$ and $S$, using
post-selected final data $t_S (x)$ on $S$.  More precisely -- and this
is an important new ingredient compared to our previous proposals --
we will use the post-selected data $t_S (x)$ for all points $x \in S$
outside the future light cone of $y$, {\it and only for those points}.

In words, our recipe is to take the expectation value for $T_{\mu \nu}
(y)$ given that the initial state was $\ket{\psi_0}$ on $S_0$,
conditioned on the measurement outcomes for $T_S (x)$ being $t_S (x)$
for $x$ outside the future light cone of $y$.  So, for any given point
$y$, our calculation ignores the outcomes $t_S (x )$ for $x$ inside
the future light cone of $y$.

Mathematically, we can calculate such expressions as follows. For any
point $y$ between $S_0$ and $S$, define the {\it effective future
  boundary} $\Lambda^1 (y)$ of $y$ in our model to be $\Lambda_1 (y )
\cup S^1 (y ) $.  Here $\Lambda_1 (y)$ is the set of points in the
lightlike future of $y$ that are either in the past of $S$ or on $S$
itself, and $S^1 (y)$ is the set of points in $S$ outside the future
light cone of $y$.  Let $ \{ S_i (y ) \}$ be a sequence of smooth
spacelike hypersurfaces that include $y$ and include all points $x
\in S^1 (y)$ such that the spacelike separation $d(x,z) \geq \epsilon_i$ for
all $z \in \Lambda_1 (y) \cap S$.  Suppose that $\epsilon_i
\rightarrow 0$ as $i \rightarrow \infty$, so that
$$
\lim_{ i \rightarrow \infty} S_i (y ) = \Lambda^1 (y) \, . 
$$
In words, the $S_i (y)$ are spacelike hypersurfaces that include
almost all
of the part of $S$ outside the future light cone of $y$, and include
$y$,
and that tend to $\Lambda^1 (y)$ as $i \rightarrow \infty$.

Now for any of the $S_i (y) $, we can consider the Born rule
probability distribution of outcomes of joint measurements of $t_S
(x)$  (for all $x \in S \cap S_i (y)$) and of $T_{\mu \nu} (y)$.  These
are calculated in the standard way, taking the initial state $\ket{
  \psi_0 }$ on $S_0$, unitarily evolving to $S_i (y)$, and applying
the measurement postulate there.  This gives us a joint probability
density function $P ( t_S (x) , t_{ \mu \nu } (y) ; x \in S \cap S_i
(y) )$.  (The notation here means that we consider the joint
proability for values of $t_S$ for all $x$ in the relevant set as 
well as for the value of 
$t_{\mu \nu}$ for the single value $y$.) By taking the limit as $i \rightarrow \infty$ we obtain a
joint probability density function $P ( t_S (x) , t_{ \mu \nu } (y) ;
x \in S^1 (y) )$.  From this, we can calculate conditional
probabilities and conditional expectations for $t_{ \mu \nu } (y)$,
conditioned on any set of outcomes for $t_S (x)$ (for 
$x \in S^1 (y)$), in the standard way.

Our mathematical description of reality, in a hypothetical
world in which physics
takes place only between $S_0$ and $S$ and in which the outcomes $t_S
(x)$ were randomly selected, is then given by the set of conditional
expectations $ \langle t_{ \mu \nu } (y) \rangle$, for each $y$
between $S_0$ and $S$, calculated as above.  We stress that the
calculations for the beables $ \langle t_{ \mu \nu } (y) \rangle$
at each point $y$ all use the same final outcome data 
$t_S (x)$.  However, different subsets of these data are used
in these calculations: for each $y$, the relevant subset is
$\{ t_S (x) \, : \, x \in S^1 (y ) \}$. 

We then consider the asymptotic limit in which $S$ tends to the
infinite future of $S_0$.   Suppose that $S_1$ is some fixed
hypersurface in the future of $S_0$. 
Let $C_{S_1}$ 
be any coarse-grained subsets of the sets of continuous
tensor functions $\{ t_{\mu \nu} (x) : x \in R^4 ,  S_0 < x < S_1 \}$,
where the notation $S_0 < x < S_1$ means that $x$ lies in the future
of some point in $S_0$ and the past of some point in $S_1$.  
Let
$$
\Prob_{S} ( C_{S_1} )
$$
be the probability that the configuration 
$\{ t_{\mu \nu} (x) : x \in R^4 ,  S_0 < x < S_1 \}$
belongs to $C_{S_1}$,
given our constructed probability density
function on the set of possible functions $T (x): S \rightarrow {\bf R} $. 
Then, assuming that 
$$
\Prob_{\infty} (C_{S_1}  ) = \lim_{S \rightarrow \infty} ( 
\Prob_{S} ( C_{S_1} ) )
$$
exists, we define this to be the probability that reality between
$S_0$ and $S_1$ is described by time-evolving mass distributions
belonging to $C_{S_1}$.  This completes this version of our proposed
description of reality.  The limiting values of $t_{\mu \nu} (x)$
define a tensor field distribution on space-time, which defines the
beables in our description.  The limiting probability distribution
defines the probability distribution on configurations of beables
(here, real tensor functions) in space-time.

\section{Toy model illustrations}

To illustrate the implications of these rules, we consider a toy
version of "semi-relativistic" quantum theory, in which a
non-relativistic system interacts with a small number of
"photons".  We treat the photons as following lightlike path
segments.  We model their interactions with the system as bounces,
which alter the trajectory of the photon.  For simplicity, we neglect
the effect of these interactions on the non-relativistic system, and
also neglect its wave function spread and self-interaction, so that in
isolation its Hamiltonian $H_{\rm sys} = 0$.  We simplify further by
working in one spatial dimension, and we take $c=1$.

{\bf Model 1:} \qquad 

We suppose that the initial state of the system is a superposition of two separate localized states, $\psi^{\rm sys}_0 =   a \psi^{\rm sys}_1 + b \psi^{\rm sys}_2$.   Here $ |a|^2 + |b|^2 = 1$ and the $\psi^{\rm sys}_i$ are states localized around the points $x = x_i$, with $x_2 > x_1$. 
For example, the $\psi^{\rm sys}_i$ could be taken to be Gaussians (but recall that we are neglecting changes in their width over time).   
We take $ | x_1 - x_2 | $ to be large compared to the regions over
which the wave functions are non-negligible.   We thus have a crude
model of a superposition of two well separated beams, or of a macroscopic
object in a superposition of two macroscopically separated states.  

We suppose that the environment consists of a single photon, initially
unentangled with the system.    It is initially propagating rightwards from the direction $x = - \infty$, so that in the absence of any interaction it would reach $x= x_1 $ 
at $t=t_1$ and $x= x_2$ at $t= t_2 = t_1 + (x_2 - x_1 )$.

We take the photon-system interaction to have the effect of 
instantaneously reversing the photon's direction of travel, while
leaving the system unaffected.   (As noted above, we neglect the
effect on the system: this violates conservation of momentum but 
simplifies the overall picture.)    

Thus, for $t<t_1$, the state of the photon-system combination in our
model is 
$$
\delta (X - x_1 - t + t_1) ( a \psi^{\rm sys}_1 ( Y ) + b \psi^{\rm sys}_2 (Y) ) \, , 
$$
where $X,Y$ are the position coordinates for the photon and system respectively.  
	
For $t_1 < t < t_2$, the state is 
$$
\delta (X - x_1 + t - t_1 ) ( a \psi^{\rm sys}_1 (Y) ) + 
\delta (X - x_1 - t + t_1 ) (  b \psi^{\rm sys}_2 (Y) ) \, . 
$$

For $ t > t_2 $, the state is 
$$\delta (X - x_1 + t - t_1 ) ( a \psi^{\rm sys}_1 (Y) ) + 
\delta (X - x_2 + t - t_2 ) (  b \psi^{\rm sys}_2 (Y) ) \, . 
$$

The possible outcomes of a (fictitious) stress-energy measurement at a
late time $t = T \gg t_2$ are thus either finding the photon heading
along the first ray $ X = x_1 + t_1 - t$ and the system localized in
the support of $\psi^{\rm sys}_1$, or finding the photon heading along
the second ray $X= x_2 + t_2 - t$ and the system localized in the
support of $\psi^{\rm sys}_2$.

Suppose, for example, we consider a real world defined by the first
outcome.  Our rules for constructing the system's beables imply that,
for $t< 2 t_1 - t_2 $, and for $x = x_1$ or $x_2$, we condition on
none of these outcomes, since all of them correspond to observations
within the future light cone.  Up to this time, then, the mass density
beables for the system are distributed according to $ | \psi^{\rm sys} (Y)
|^2$, with a proportion $ | a |^2 $ localized around $Y= x_1$ and a
proportion $ | b |^2 $ localized around $Y= x_2$.

For $ 2 t_1 - t_2 < t < t_1 $, the observation of the photon on the first ray is outside the future light cone of the component of the system localized at $x_2$, but not outside the future light cone of the component localized at $x_1$.    This gives us mass density beables distributing a proportion $ | a |^2 $ of the total system mass around $x_1$, but zero mass density beables around $x_2$. 

For $ t > t_1 $, the observation is outside the future light cone of both localized components of the system.    This gives us mass density beables distributing the full system mass around $x_1$, and zero around $x_2$. 

In other words, in the picture given by the beables, the system is a
combination of two mass clouds with appropriate Born rule
weights initially, and ``collapses" to a single cloud containing the full mass after $ t > t_1$. 

In the frame defined by our coordinates, this picture has the
seemingly peculiar property that, for $ 2 t_1 - t_2 < t < t_1 $, the
object is, so to speak, only "partly present", with only a proportion
$|a|^2$ of its mass represented by the beables.  The point here is
that the "realization" of the object is determined in this case by the
reflected light ray corresponding to the photon wave function $$\delta
(X - x_1 + t - t_1 ) $$.  Along lines parallel to this light ray, the
system collapses from a combination of two mass clouds to a single
cloud of full mass, without any intervening interval in which "the
mass is only partly present".  In general, we expect our rules to
define "collapses" in a way that will 
depend on the relationship between the frame considered and the
possible interactions between the particles.  However, with 
realistic interaction rules and initial states it will not generally
define ``collapses'' to be instantaneous and need not necessarily
associate the same frame with all collapses.   

Note that, in this model, so long as $T \gg t_2$, its precise value
makes no difference to the conclusions set out so far.  To this
extent, our asymptotic condition holds.

To complete the story in this model, we should characterize the
beables associated with the photon.  For $t<t_1$, this is
unproblematic: the photon follows the incoming light ray $\delta (X -
x_1 - t + t_1)$.  It also follows from our rules that, given a final
detection of the photon along the outgoing light ray $\delta (X - x_1
+ t - t_1 )$, the second possible outgoing light ray $\delta (X - x_2
+ t - t_2 )$ is empty of beables, since the observation of the first
light ray takes place outside the future light cone of points on the
second.  It would be tempting, then, to say that the beable description
implies that the photon follows the first outgoing light ray.
However, our rules imply a discontinuity precisely along this light
ray.  The only relevant measurement outcome (until $t$ close to $T$)
is the observation of the photon at the end of the light ray, and this
is lightlike rather than spacelike separated from points along the
ray.  Strictly speaking, given that we consider the limit of spacelike
surfaces approaching the ray, this observation never enters into the
calculation, so that the beables describe only $|a|^2$ of the photon
energy-momentum following this ray.  The fact that other final state
observations (of the system, and of the absence of the photon from the
second possible outgoing light ray) become spacelike separated for $t$
close to $T$ does not rescue this part of the picture, since we
ultimately take the limit as $T \rightarrow \infty$.  We tentatively
interpret this feature as an artefact, due to the parsimony of this
model, since if many photons were interacting with the system, from
both directions, we expect some final measurement outcomes to be
spacelike separated from all possible outgoing rays.   This point
is illustrated in our next model.  

{\bf Model 2:} \qquad 

To test these intuitions a little further, we consider a symmetric two
photon version of the above model.  The initial system wave function
is as before.  Now, though, there are two incoming photons, one from
the left, which arrives at $x_1$ at time $t_1$, and the other from the
right, which arries at $x_2$ at the same time $t_1$.  As before, when
interacting with the system, each photon bounces, instantaneously
reversing its direction of travel, while the system is unaffected.

Thus, for $t<t_1$, the state of the photon-system combination is 
$$
\delta (X_1 - x_1 - t + t_1) \delta (X_2 - x_2 + t - t_1 ) 
( a \psi^{\rm sys}_1 ( Y ) + b \psi^{\rm sys}_2 (Y) ) \, , 
$$
where $X_1 , X_2$ are the position coordinates for the two photons and $Y$ is the coordinate of the system.  

Suppose, again, that we find a final measurement outcome consistent
with the two photons bouncing off the system localized at $x_1$.  That
is, at late time $T$ we find one photon along the intersection of
$t=T$ with the ray $\delta (X_1 - x_1 + t - t_1 )$, a second at the
intersection of $t=T$ with the ray $\delta (X_2 - x_1 - t + t_2 )$,
and the system localized at $x_1$.

Now there are four relevant outgoing rays: the two at whose ends we
find the two photons, the ray $\delta (X_1 - x_2 + t - t_2 )$, at
whose end we do not find photon $1$, and the ray $\delta (X_2 - x_2 -
t + t_1 )$, at whose end we do not find photon $2$.  These last two
rays correspond to the reflection of a photon from the system when
localized at $x_2$.

These extra rays change the picture.   For the system, our rules now give beables that distribute its mass density in the expected proportions 
( $| a |^2$ and $ | b |^2 $) up to time $t= t_1 - (x_2 - x_1)$. 
At this point, the beables at both $x_1 $ and $x_2$ "collapse'', and from then onwards all the beable mass density is localized at $x_1$, with none at $x_2$.    

As for the photons, the beables corresponding to their incoming states
show them along the incoming rays, as expected, until $t= t_1$.
Because the measurement outcome corresponding to the absence of a
photon along the outgoing ray $\delta (X_2 - x_2 - t + t_1 )$ is
outside the future light cone of the outgoing ray $\delta (X_1 - x_1 +
t - t_1 ) $, and vice versa, the beables describe a full photon along
the latter ray and none along the former.  The outcomes for these two
rays also ensure that the beables describe a full photon along the ray
$\delta (X_2 - x_1 - t + t_2 )$ and none along $\delta (X_1 - x_2 + t
- t_2 )$.

The entire beable picture for system and photons is thus consistent
(if possibly initially a little challenging to the intuition).
According to the beables, the system collapses to a mass density cloud
localized around $x=x_1$ at time $t= t_1 - x_2 + x_1$.  The photons
follow definite paths, propagating until they hit this localized
cloud, and then bouncing to reverse their courses.

The symmetry of this example ensures that this picture would be
simply reflected if we obtained the other set of possible measurement 
outcomes at late time $T$,
with the system beables now localizing around $x=x_2$ and the photons
now bouncing off this localized cloud.

\section{Discussion}  

We have presented a new proposal for a solution to the Lorentzian
quantum reality problem.  The proposal has several features in common
with earlier suggestions \cite{akrealworld,akreality}.  But it has a 
significant new feature: beables at any point in space-time are
defined using only final outcome data from outside that point's future
light cone.  We are particularly interested in the beables 
associated with massive quasiclassical systems -- that is, physical
systems whose coarse-grained mass densities follow approximately
classical equations of motion most of the time, with stochastic 
corrections arising (inter alia) from interactions with microscopic quantum
systems \cite{gell1990quantum}.   
The intuition here is that such a quasiclassical system
generally leaves effective records of its
location outside its future light cone, arising from 
its possible interactions with photons and
other massless particles.  This means that we could
infer beables characterising its quasiclassical behaviour at any given
time from a (hypothetical) measurement at late times, even if we
restrict attention to measurement data outside its future light cone,
since these data include records arising from recent interactions with
photons.   Following this prescription allows us to calculate a
beable picture by an algorithm that involves only calculating the
outcome probabilities for various measurements in standard (unitary)
relativistic quantum theory.  

Our proposal was set in the context of Minkowski space-time, but can
be simply extended to other background space-times that have
well-defined asymptotic futures.  We do 
not believe it is true in all cosmological models that quasiclassical
systems leave effective records outside their future light cone.  
For example, models with a final ``big crunch'' do not have 
this property.   
In this sense, our proposal is contingent on both the large-scale
structure of space-time and the particular quantum dynamics defining
the quantum evolution law within that space-time.  However, 
while future cosmological observations that modify the presently
preferred cosmological models could possibly change the picture, we do not
presently see any compelling reason to believe that it is not 
viable in our own universe.   At present, it seems that photons and
(if they exist) gravitons and 
other massless particles interacting with quasiclassical structures
in our cosmos can and apparently do escape to future infinity, 
without further interactions that significantly alter their direction
or slow their progress, at an appreciable rate. Indeed, the observed
accelerating expansion of the universe seems (if it continues 
indefinitely) to guarantee this will always be the case. 

We have illustrated our proposal in toy models, and shown that it
gives sensible and intuitively appealing descriptions of 
quantum reality in these models.   The models considered here
are admittedly extremely simplified.   
They are set in one space and one time dimension.
While the choice of one space dimension makes the geometry
simple, it introduces an unrealistic feature.   If we
considered many massive systems localized in different places, and 
many incoming photons, and also maintained the simplified 
interaction rule that photons always bounce from massive
systems, then photons would always remain in the regions to
the left or right of all the massive systems, or else
trapped between two systems.   This is an artefact of the
$1+1$ dimensional geometry and of the unrealistic assumption
that interactions between photons and matter produce a bounce
with probability one.   In a $3+1$ dimensional geometry
in which the angular distribution of massive objects has large
gaps and in which photons are incoming from all directions
(a more realistic model of the present state of our
own cosmos), and with realistic interactions,
photons scattering from  a massive object
can and often will escape to future infinity without further
scatterings that significantly alter their space-time path. 

Our toy models also neglect Schr\"odinger
dynamics for massive particles or objects, setting $H=0$ for
these subsystems.  They treat photons and particles as effectively
pointlike, rather than allowing for some spread in their wavepackets,
and assume a simplified interaction in which the photons undergo
perfect instantaneous reflection without causing any reaction on
the massive reflecting particles.  We believe they nonetheless
illustrate the key intuitions well, and indicate that more realistic 
$3+1$ dimensional
models that allow for spreading wavepackets and imperfect interaction
with reaction would produce beables giving the same qualitative description of 
reality.   Of course,  demonstrations in more sophisticated models would
be useful: we aim to produce such demonstrations in future works.  
  
Our models also necessarily rely on a hybrid ``semi-relativistic''
treatment of photons travelling at light speed interacting with
particles or objects for which relativistic effects are negligible.
This treatment can certainly be significantly improved, 
although it will be hard to provide a fundamentally correct and rigorous
treatment, given the difficulties in modelling any realistic
quasiclassical physics within quantum field theory as presently
understood.   
Nonetheless, the consistency of quantum field theory with
Minkowski causality, reflected in the fact that 
propagators fall off rapidly outside the future light cone
and in the existence of the Tomonaga-Schwinger formalism, suggests to us
that the strategy of defining beables at a point by conditioning on
final outcomes outside its future light-cone is natural in 
quantum field theory.   It suggests too that the qualitative
features of the beables descriptions of reality in our toy
models should accurately reflect the pictures that should
emerge from a fully relativistic field-theoretic treatment.       

Mass density beable ontologies were first proposed 
for non-relativistic collapse models by Pearle and Squires
(\cite{PhysRevLett.73.1}; see also \cite{ghirardi1995describing}).  
An extension of these ontologies to relativistic collapse models using 
constructions previously defined   
in \cite{kent2005nonlinearity,akcausalqt}
was proposed in \cite{bedingham2011relativistic}.  
These proposals apply to generalizations of quantum theory
rather than to quantum theory itself.  

In contrast, the solution proposed here requires no alteration
to quantum dynamics.   Our models predict ``collapses'' in 
the sense that the beables are initially consistent with a superposition
of macroscopically distinct states, and later with a single component
of the superposition.   However, these ``collapses'' have no direct effect
on the underlying dynamics: rather, they are all inferred from the
final outcome state.   In particular, unlike non-relativistic or
relativistic dynamical collapse 
models \cite{ghirardi1986unified,GPR,tumulka2006relativistic,
bedingham2011relativistic}, our models predict no violation
of conservation of energy.   

Our proposal is also consistent with the standard understanding of
Bell experiments.   Without producing a detailed toy model here, 
it is easy to give the basic intuitions. 
First, it predicts that the beables associated with quasiclassical measuring
devices in two spacelike separated wings of a Bell experiment 
will correspond to definite measurement outcomes after the experiment,
since the final state will include ``photon'' states characterising
one of the possible outcomes on each wing.  
Second, it predicts that neither the beables in the entangled pair of
particles being measured, nor those in the measuring devices,
are correlated with the eventual outcomes before the measurements.    
No ``photons'' significantly interact with the entangled particles
before the measurements, and the ``photon'' interactions with the
measuring devices are the same whatever the eventual outcomes.
Third, it predicts the same correlations among measurement outcomes
as those predicted by standard quantum theory, since the dynamics
are those of standard quantum theory, and its predictions for 
measurement outcomes are inferred from the final state records 
of those outcomes, whose probability distribution is obtained
from standard quantum dynamics and the Born rule.   

Our solution also seems to fit naturally within 
the existing framework for relativistic quantum field theory
and to be consistent with standard intuitions about relativistic
quantum field theories.   Of course, making it mathematically
rigorous nonetheless seems impossible without solving the 
problem of rigorously defining realistic relativistic quantum 
field theories in $3+1$-dimensional Minkowski space.
Perhaps for the moment we should be content to have a coherent
conceptual account of reality defined at the same (low) level of 
rigour as our understanding of relativistic quantum field theory. 
In this context as elsewhere, the programme of 
rigorizing existing relativistic quantum field theories may not necessarily 
be the most scientifically fruitful way forward: other options
include investigating quantum theories of gravity or other potentially
deeper underlying theories.     

The proposal presented here admits many variations, 
including the various 
interesting possibilities discussed in Ref. \cite{akreality}
(adapted to allow for the new treatment here of outcome data,
distinguishing data from inside and
outside the future lightcone). 
Intriguingly, it also suggests
natural ways to generalize quantum theory, following
\cite{kent1998beyond,kent2013beable,akreality}.   For, given a 
beable formulation -- a description of quantum reality --
consistent with standard quantum dynamics, we can define
modified theories by altering the rules for assigning
probabilities to beable configurations.   For example,
we can introduce weight factors that depend directly on the 
beable configuration, and use these to rescale the probabilities
for beable configurations that are defined by standard quantum
theory.   Even more general possibilities are also available.

More extended comments and
discussions of possibilities in these directions, which apply --
and indeed, apply with more force -- to the present proposal, can be found in
\cite{akreality}.
The fact that our beables take the form of a generalized stress-energy
tensor also suggests a possible new path to considering the
coupling of quantum theory and gravity, while avoiding the problems
afflicting standard semi-classical gravity \cite{akreality}.

\section{Acknowledgements}
This work was partially supported  by Perimeter Institute
for Theoretical Physics. Research at Perimeter Institute is supported
by the Government of Canada through Industry Canada and by the
Province of Ontario through the Ministry of Research and Innovation.

\section*{References}

\bibliographystyle{plain}
\bibliography{reality}{}

\end{document}